# Single crystal growth and magnetism of $Sr_3NaIrO_6$ and $Sr_3AgIrO_6$: tracking the $J = 0$ ground state of $Ir^{5+}$

Peng-Bo Song (宋鹏博)[1,2], Zhiwei Hu[3], Su-Yang Hsu[4], Jin-Ming Chen[4], Jyh-Fu Lee[4], Shan-Shan Miao (苗杉杉)[1], You-Guo Shi (石友国)[1,2,5]*, Hai L. Feng (冯海)[1,5]*

[1]*Beijing National Laboratory for Condensed Matter Physics and Institute of Physics, Chinese Academy of Sciences, Beijing 100190, China*

[2]*Center of Materials Science and Optoelectronics Engineering, University of Chinese Academy of Sciences, Beijing 100190, China*

[3]*Max Planck Institute for Chemical Physics of Solids, Dresden 01187, Germany*

[4]*National Synchrotron Radiation Research Center, Hsinchu 30076, Taiwan, Republic of China*

[5]*Songshan Lake Materials Laboratory, Dongguan, Guangdong 523808, China*

*Corresponding authors:

ygshi@iphy.ac.cn (YGS)

hai.feng@iphy.ac.cn (HLF)

**Abstract.** Single crystals of $Sr_3NaIrO_6$ and $Sr_3AgIrO_6$ have been successfully grown using hydroxides flux. $Sr_3NaIrO_6$ and $Sr_3AgIrO_6$ crystallize in the $K_4CdCl_6$-type structure with the space group $R$-3$c$ (No.167). $Sr_3NaIrO_6$ and $Sr_3AgIrO_6$ are electrically insulating with estimated activation gaps of 0.68 eV and 0.80 eV, respectively. $Sr_3NaIrO_6$ and $Sr_3AgIrO_6$ show paramagnetic behavior down to 2 K. In this work, the observed effective moments, $\mu_{eff}$, for $Sr_3NaIrO_6$ single crystals are 0.31 $\mu_B$ for $H \perp c$ and 0.28 $\mu_B$ for $H \| c$, which are much smaller than that of 0.49 $\mu_B$ previously reported for the polycrystalline $Sr_3NaIrO_6$ samples. For $Sr_3AgIrO_6$ single crystals, a much larger value of $\mu_{eff}$ = 0.57 $\mu_B$ is observed as compared with $Sr_3NaIrO_6$ single crystals. The x-ray absorption spectroscopy and low-temperature specific heat data indicate that the Ir in $Sr_3NaIrO_6$ has an almost pure $Ir^{5+}$ valence state, while the Ir in $Sr_3AgIrO_6$ is slightly lower than +5. The estimated low limits of magnetic impurity $Ir^{4+}$ are about ~1.7% and ~9.2% for $Sr_3NaIrO_6$ and $Sr_3AgIrO_6$, respectively. These magnetic impurities are likely to fully explain the observed $\mu_{eff}$ values for $Sr_3NaIrO_6$ and $Sr_3AgIrO_6$ single crystals, supporting the $J = 0$ ground state of $Ir^{5+}$ in $Sr_3NaIrO_6$ and $Sr_3AgIrO_6$.

## Introduction

In 3d transition metal oxides, the valence electrons are strongly correlated and the Hubbard $U$ plays an important role [1–5]. In comparison with 3d electrons, the orbitals of 5d electrons are more extended and the $U$ in the 5d system becomes weaker while the spin-orbit coupling (SOC) becomes much stronger due to their larger atomic number [6,7]. In the strong SOC regime, the SOC can split three $t_{2g}$ orbitals in the octahedral crystal field into an upper $j = 1/2$ doublet and a lower $j = 3/2$ quadruplet [8,9]. For example, in tetravalent iridate $Sr_2IrO_4$ ($Ir^{4+}$: $5d^5$) the SOC-assisted Mott-insulating state is explained with the $J = 1/2$ ground state [9]. Resonant inelastic x-ray scattering measurements on pentavalent osmates ($Os^{5+}$: $5d^3$) reveal the SOC controlled $J = 3/2$ ground state [10]. In such a scenario, if there are four 5d electrons filling the lower quadruplet, the ground state should be $j = 0$. Long-range magnetic orders reported in $Ir^{5+}$ ($5d^4$) double perovskite oxides $Sr_2YIrO_6$ and $Ba_2YIrO_6$ with effective moment ($\mu_{eff}$) of 0.91 $\mu_B$/Ir and 1.44 $\mu_B$/Ir, respectively, raise concerns about the ground state of $5d^4$ oxides [11,12]. These results have been challenged by other studies reporting that no magnetic order was found in $Ba_2YIrO_6$ [13] and $Sr_2YIrO_6$ [14] down to ~430 mK. Studies on $A_2YIrO_6$ (A = Sr, Ba) and other $Ir^{5+}$ double perovskite oxides generally reveal a weak paramagnetic behavior with small $\mu_{eff}$ values of 0.19 – 0.63 $\mu_B$/Ir [13–25] which are much lower than the theoretical spin-only $\mu_{eff}$ = 2.83 $\mu_B$/Ir demonstrating a SOC dominated ground state. The origin of these finite magnetic moments is still ambiguous. Quench of the $J = 0$ state for $Ir^{5+}$ due to $IrO_6$ octahedra distortion in $Sr_2YIrO_6$ was proposed by Cao et al [11]. However, this scenario cannot explain the paramagnetic moment observed in cubic $Ba_2YIrO_6$ where is no structural distortion, and the studies on $Ba_{2-x}Sr_xYIrO_6$ studies do not find correlations between $\mu_{eff}$ values and the degree of structural distortions [19,26]. The existence of magnetic impurities has been suggested by studies in $Sr_2YIrO_6$ [14] and $Ba_2YIrO_6$ [17]. Fuchs et al. confirmed the existence of $Ir^{4+}$ and $Ir^{6+}$ magnetic defects which are responsible for the magnetism in $Ba_2YIrO_6$ [18]. The antisite disorder in double perovskites has also been suggested to play an important role [16,20]. Laguna-Marco et al suggest that the $Ir^{4+}$ and $Ir^{6+}$ magnetic impurities may locate in the antisite disorder region [20]. Condensation of $J = 1$ triplon excitations of $5d^4$ oxides is also a possible source for the observed magnetic moments [27,28]. Chen et al. proposed that the condensation is unlikely in $Sr_2YIrO_6$ and $Ba_2YIrO_6$ with the ideal crystal structure, but the antisite disorder between $Y^{3+}$ and $Ir^{5+}$ can break down the local nonmagnetic singlets [16]. Recent studies on $A_2BIrO_6$ ($A$ = Ba, Sr; $B$ = Lu, Sc) also support the $J = 0$ ground state for these $Ir^{5+}$ oxides and indicate the magnetic signals are from extrinsic sources, such as magnetic impurities and antisite disorder [21].

To narrow down the possible explanations, it is better to studies on $Ir^{5+}$ oxides with less antisite disorder. Recently, studies on $K_4CdCl_6$-type polycrystalline $Ir^{5+}$ oxide $Sr_3NaIrO_6$ have been reported and indicate a possible quantum spin liquid ground state (reported $\mu_{\mathrm{eff}} = 0.49\ \mu_{\mathrm{B}}/\mathrm{Ir}$) [22]. In comparison with $A_2YIrO_6$ (A = Sr, Ba), where the $Ir^{5+}O_6$ octahedra are separated by $Y^{3+}O_6$, the $Ir^{5+}O_6$ octahedra are separated with $Na^{1+}O_6$ octahedra in the $Sr_3NaIrO_6$. The larger charge difference would reduce the antisite disorder between $Na^{1+}$ and $Ir^{5+}$ in $Sr_3NaIrO_6$ as compared with $Sr_2YIrO_6$ and $Ba_2YIrO_6$. Thus, the $K_4CdCl_6$-type iridate is a good platform to investigate the ground state of $Ir^{5+}$ ions. To track the $J = 0$ gound state for $Sr_3NaIrO_6$, it is better to grow single crystals to minimize any by-phases and lattice defects. In this work, we successfully grow single crystals of $K_4CdCl_6$-type iridate oxides $Sr_3NaIrO_6$ and $Sr_3AgIrO_6$. Magnetic measurements reveal that the $\mu_{eff}$ for $Sr_3NaIrO_6$ single crystals is about 0.31 $\mu_B$ for $H \perp c$ and 0.28 $\mu_B$ for $H \| c$ which are smaller than that of 0.49 $\mu_{\mathrm{B}}$ reported for the polycrystalline $Sr_3NaIrO_6$ [22]. The presence of a few percent of magnetic $Ir^{4+}$ impurity is indicated by the analysis of low-temperature specific heat data which is likely to fully explain the observed $\mu_{eff}$, supporting the $J = 0$ ground state of $Ir^{5+}$ in $Sr_3NaIrO_6$.

## Experiment

Single crystal samples of $Sr_3NaIrO_6$ and $Sr_3AgIrO_6$ were both prepared by flux method using NaOH-KOH and KOH, respectively. The $SrOH \cdot 8H_2O$, NaOH, KOH, and Ir with a molar ratio of 3:50:40:1 (for $Sr_3NaIrO_6$) and $SrOH \cdot 8H_2O$, $Ag_2O$, KOH, and Ir with a molar ratio of 3:0.5:40:1 (for $Sr_3AgIrO_6$) were weighted, respectively. The mixtures were placed into $Al_2O_3$ crucibles with lids, then heated to 873 K in 1 h and annealed for 12h before cooled to room temperature by stopping the heating. Single crystals were separated by washing with deionized water.

Single-crystal x-ray diffraction measurements were conducted on a Bruker D8 Venture diffractometer at 300K using Mo Kα radiation (λ = 0.71073 Å). The frames were integrated with the Bruker SAINT software package using a narrow-frame algorithm. Data were corrected for absorption effects using the multi-scan method (SADABS). The crystalline structures were refined by the full-matrix least-squares method on $F^2$ using the SHELXL-2018/3 program.

Single crystals of selected samples were used for magnetic susceptibility (χ), longitudinal resistivity $\rho$, specific heat, and x-ray absorption spectroscopy (XAS). The magnetic properties were measured under different applied magnetic fields in Field-Cooling (FC) and Zero-Field-Cooling (ZFC) modes using SQUID-VSM device in a magnetic properties measurement system (MPMS). Isothermal magnetization (M-H) was measured at several fixed temperatures. These transport measurements (both resistivity and specific heat) were measured by a physical property measurement system (Quantum Design inc) using the standard DC four-probe technique and a thermal relaxation method, respectively. XAS spectra at the Ir-$L_3$ edges were studied at the Taiwan Light Source (TLS) beamline 17C of the National Synchrotron Radiation Research Center (NSRRC).

## RESULTS AND DISCUSSION

Single crystals for $Sr_3NaIrO_6$ with dimensions of ~2*mm*×0.3*mm*×0.3*mm* and Single crystals for $Sr_3AgIrO_6$ with ~0.3*mm*×0.3*mm*×0.3*mm* were obtained as shown in Figs. 1(c) and 1(d), respectively. Analysis of the room temperature SCXRD data confirms that $Sr_3NaIrO_6$ and $Sr_3AgIrO_6$ crystallize in the $K_4CdCl_6$-type structure with the space group *R*-3*c* (No.167). The refined lattice parameters were a = 9.6408(3) Å, and c = 11.5508(5) Å for $Sr_3NaIrO_6$ and a = 9.5996(3) Å, and c = 11.9032(6) Å for $Sr_3AgIrO_6$. Detailed crystallographic data obtained from the SCXRD were summarized in Table I. The crystallographic information files for $Sr_3NaIrO_6$ and $Sr_3AgIrO_6$ are attached in the Supplemental Material [29].

Table I. Crystallographic and structure refinement data for $Sr_3AgIrO_6$ and $Sr_3NaIrO_6$.

| Chemical formula | $Sr_3AgIrO_6$ | $Sr_3NaIrO_6$ |
|---|---|---|
| Formula weight | 658.94 g/mol | 574.05 g/mol |
| Radiation | Mo K$\alpha$, 0.71073 Å | Mo K$\alpha$, 0.71073 Å |
| Temperature | 300 K | 300 K |
| Crystal system | Trigonal | Trigonal |
| Space group | *R*-3*c* | *R*-3*c* |
| Unit cell dimensions | a = 9.5996(3) Å<br>c = 11.9032(6) Å | a = 9.6408(3) Å<br>c = 11.5508(5) Å |
| Volume | 949.95(8) Å$^3$ | 929.76(7) Å$^3$ |
| Z | 6 | 6 |
| Density(calculated) | 6.911 g/cm$^3$ | 6.152 g/cm$^3$ |
| Absorption coefficient | 49.048 mm$^{-1}$ | 47.117 mm$^{-1}$ |
| No. of reflections | 5112 | 6376 |
| No. independent reflections | 265 | 261 |
| No. observed reflections | 256 | 256 |
| F (000) | 1716 | 1500 |
| Theta range for data collection | 4.21 to 28.29° | 4.23° to 28.36° |
| Index ranges | $-12 \leq h \leq 12$<br>$-12 \leq k \leq 12$<br>$-15 \leq l \leq 15$ | $-12 \leq h \leq 12$<br>$-12 \leq k \leq 11$<br>$-15 \leq l \leq 15$ |
| Goodness of fit | 1.265 | 1.118 |
| R1 ($I > 2\sigma_1$) | 0.0230 | 0.0103 |
| ωR2($I > 2\sigma_1$) | 0.0735 | 0.0258 |
| R1 (all data) | 0.0237 | 0.0106 |
| ωR2(all data) | 0.0740 | 0.0261 |
| Weighting scheme | $w = 1/[\sigma^2 F_o^2 + (0.0423P)^2 + 38.8002P]$ Where $P = (F_o^2+2F_c^2)/3$ | $w = 1/[\sigma^2 F_o^2+(0.0141P)^2+ 6.8962P]$ where $P=(F_o^2+2F_c^2)/3$ |
| Refinement software | SHELXL-2018/3 | SHELXL-2018/3 |

Table II. Refined atomic positions and temperature parameters for $Sr_3NaIrO_6$.

| atom | *x* | *y* | *z* | Occupancy | $U_{eq}$ (Å$^2$) | Site |
|---|---|---|---|---|---|---|
| Ir1 | 2/3 | 1/3 | 5/6 | 1 | 0.00379(9) | 6b |
| O1 | 0.4882(2) | 0.3102(2) | 0.7336(2) | 1 | 0.0075(4) | 36f |
| Sr1 | 0.3090(1) | 1/3 | 7/12 | 1 | 0.0069(1) | 18e |
| Na1 | 2/3 | 1/3 | 7/12 | 1 | 0.0090(6) | 6a |

Table III. Refined atomic positions and temperature parameters for $Sr_3AgIrO_6$.

| atom | *x* | *y* | *z* | Occupancy | $U_{eq}$ (Å$^2$) | Site |
|---|---|---|---|---|---|---|
| Ir1 | 2/3 | 1/3 | 1/3 | 1 | 0.0061(3) | 6b |
| O1 | 0.6896(7) | 0.5138(6) | 0.2373(4) | 1 | 0.0099(10) | 36f |
| Sr1 | 0.6896(1) | 2/3 | 5/12 | 1 | 0.0094(3) | 18e |
| Ag1 | 2/3 | 1/3 | 1/12 | 1 | 0.0230(5) | 6a |

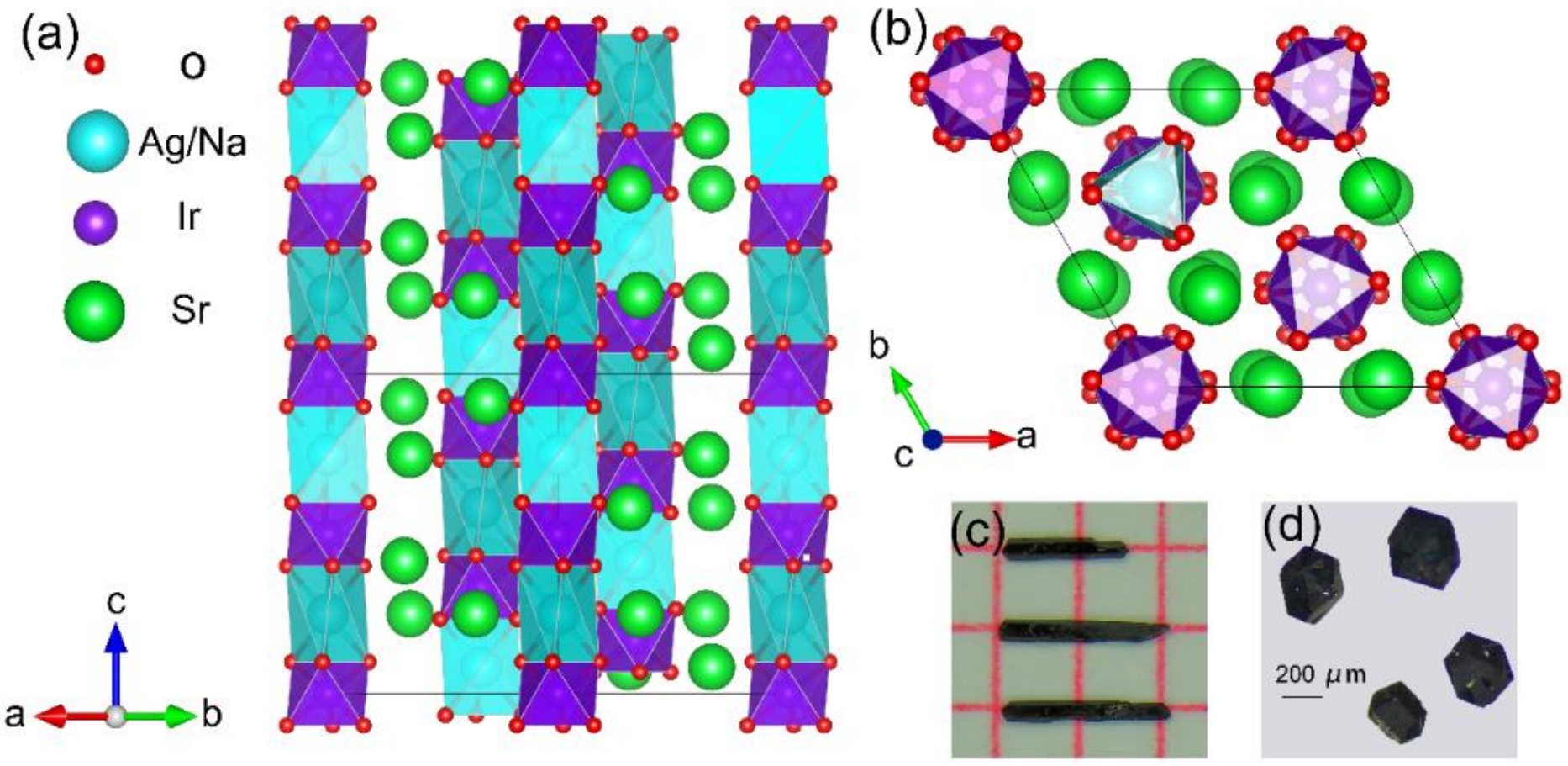

Figure 1. Crystal structures of $Sr_3AgIrO_6$($Sr_3NaIrO_6$) view along (a) [110] and (b) [001] directions. (c) and (d) shows optical photos of $Sr_3NaIrO_6$ and $Sr_3AgIrO_6$ single crystals, respectively.

In the crystal structures of $Sr_3NaIrO_6$ and $Sr_3AgIrO_6$, Na/Ag atoms occupy the 6*a* site, Ir atoms occupy the 6b site, Sr atoms occupy the 18e site, and O atoms occupy the 36f site (see Table II and Table III). The crystal structure of $Sr_3NaIrO_6$ and $Sr_3AgIrO$ is shown in Figure 1. The $IrO_6$ octahedra are connected with $NaO_6$/$AgO_6$ octahedra through face-sharing, forming one-dimensional chains. During the analysis of single-crystal x-ray diffraction data, we checked the possibility of antisite between Na/Ag and Ir sites and found no indication of antisite disorder. The bond length for Ir–O are 1.984 Å and 1.989 Å for $Sr_3NaIrO_6$ and $Sr_3AgIrO_6$, respectively. The bond valence sum for Ir ions calculated from the Ir-O bond length is 4.99 and 4.92 for $Sr_3NaIrO_6$ and $Sr_3AgIrO_6$, respectively, which are close to the nominal $Ir^{5+}$ oxidation states.

It is well known that hard x-ray absorption spectroscopy (XAS) at the 5d elements L edge is highly sensitive to their oxidation states, since the energy position of the strong white line shifts to higher energy by one or more eV if an increase of the valence state of 5d metal ion by one [30–34]. Figure 2 shows the Ir-$L_3$ XAS spectra for $Sr_3NaIrO_6$ and $Sr_3AgIrO_6$ together with the $Ir^{4+}$ reference $La_2CoIrO_6$ and the $Ir^{5+}$ reference $Sr_2CoIrO_6$ [35]. The energy position of $Sr_3NaIrO_6$ shifts about ~1.3 eV toward higher energy in comparison with $La_2CoIrO_6$, but locates at the same energy as that of $Sr_2CoIrO_6$, supporting the $Ir^{5+}$ oxidation state. The energy position of $Sr_3AgIrO_6$ is a little lower than those of $Sr_3NaIrO_6$ and $Sr_2CoIrO_6$, and is about ~1 eV higher than $La_2CoIrO_6$, indicating that the oxidation state of Ir ion in $Sr_3AgIrO_6$ is little lower than +5. From these XAS spectra we can conclude that the oxidation state of Ir ion in $Sr_3NaIrO_6$ are $Ir^{5+}$, but we cannot exclude the presence of a few percent of $Ir^{4+}$ or $Ir^{6+}$ ions [35]. For $Sr_3AgIrO_6$, the average Ir oxidation state is little lower than $Ir^{5+}$ and a moderate amount of $Ir^{4+}$ ions is coexisting with the major $Ir^{5+}$ ions.

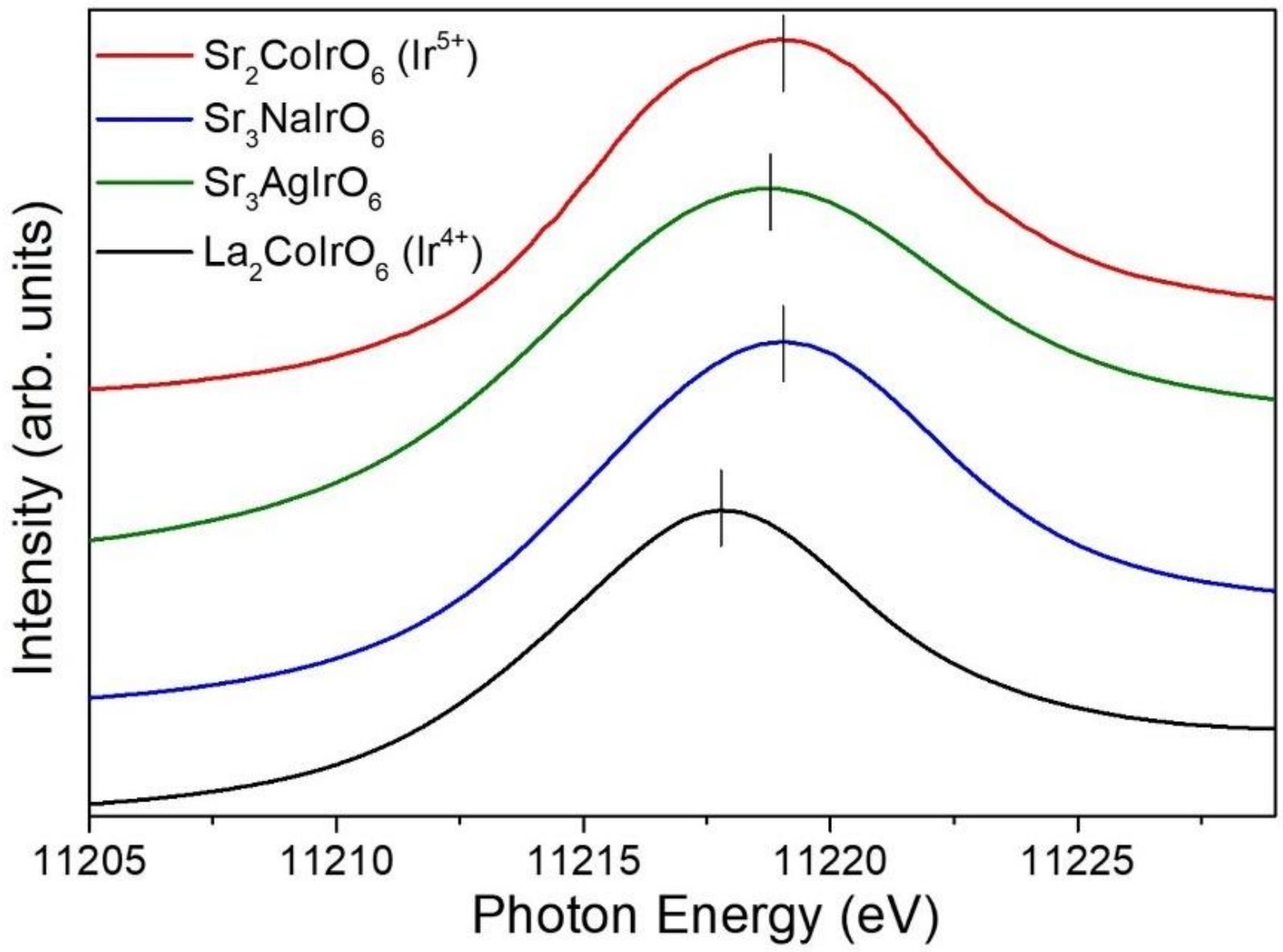

Figure 2. Ir-$L_3$ XAS spectra of $Sr_3NaIrO_6$ and $Sr_3AgIrO_6$ in comparison with an $Ir^{4+}$ reference $La_2CoIrO_6$ and an $Ir^{5+}$ reference $Sr_2CoIrO_6$.

The temperature-dependent resistivity, $\rho(T)$, for $Sr_3NaIrO_6$ and $Sr_3AgIrO_6$ are shown in Figs. 3(a) and 3(b). The $\rho(T)$ curves show semiconducting behavior. The $\rho$ increases with cooling and is out of range below 250 K and 260 K for $Sr_3NaIrO_6$ and $Sr_3AgIrO_6$, respectively. The high-temperature data (>320 $K$) are used to estimate the gap according to the Arrhenius equation, $\rho \propto exp^{(E_a/2k_BT)}$ [see the insets in Figs. 3(a) and 3(b)]. The estimated activation gap, $E_a$, is ~0.80 eV for $Sr_3NaIrO_6$ and ~0.58 eV for $Sr_3AgIrO_6$.

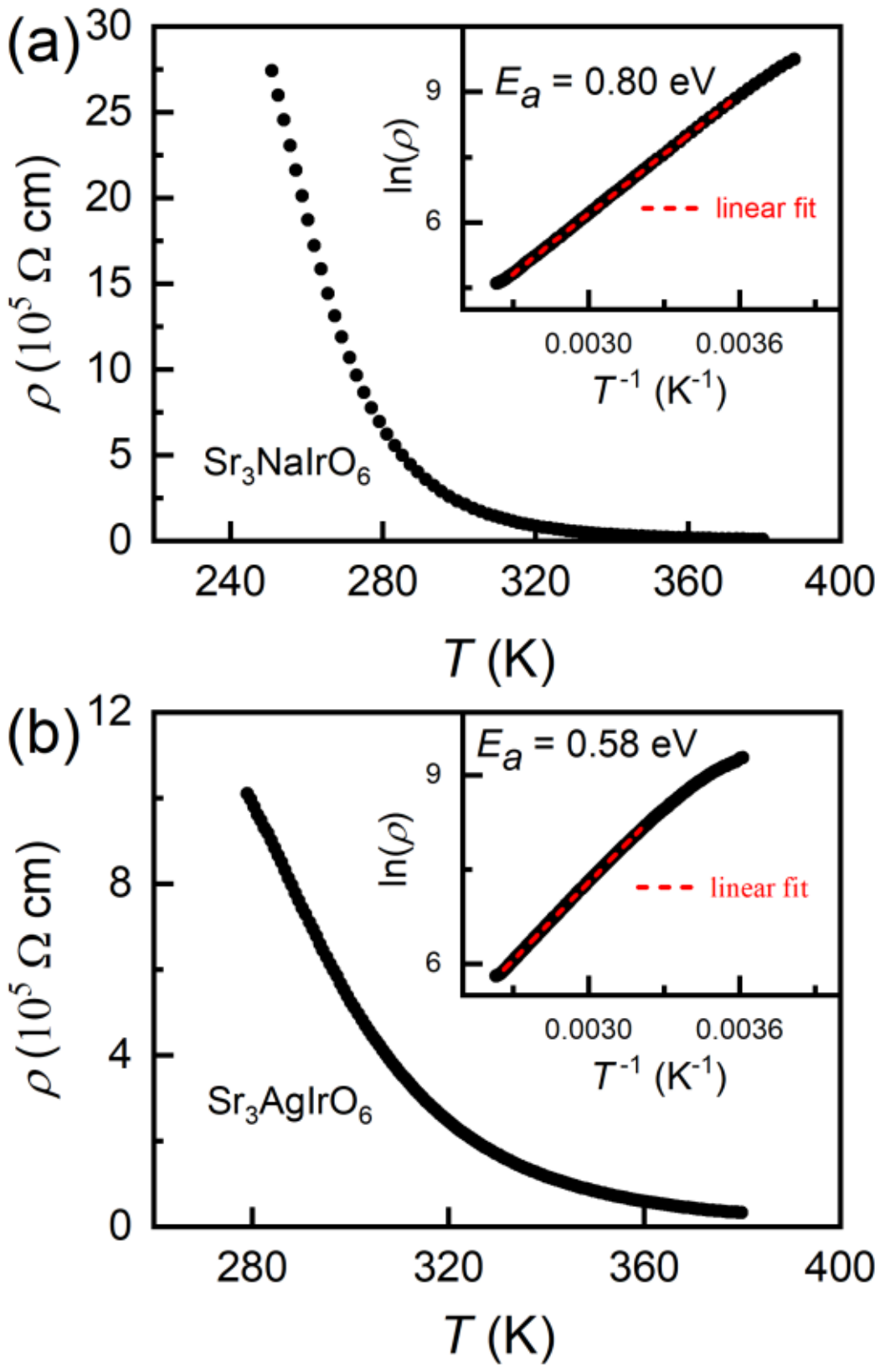

Figure 3. The temperature dependence of resistivity for (a) $Sr_3NaIrO_6$ and (b) $Sr_3AgIrO_6$. The insets show corresponding data fitted with the Arrhenius equation.

Single crystals of $Sr_3NaIrO_6$ are large enough (~2 mm along the *c*-axis) to measure their anisotropic magnetic properties. The temperature-dependent magnetic susceptibility curves, $\chi(T)$, of $Sr_3NaIrO_6$ single crystals measured with magnetic fields perpendicular to the *c*-axis ($H\perp c$) and parallel to the *c*-axis ($H||c$) are shown in Figs. 4(a) and 4(b). The ZFC and FC curves are overlapping, and only FC curves are shown. There is no sign of magnetic order down to 2 K. The $\chi$(T) data above 50 K can be fitted with the Curie-Weiss law, $\chi = \chi_0 + C/(T - \theta_W)$, where $C$, $\theta_W$, and $\chi_0$ are the Curie constant, Weiss temperature, and the temperature-independent component, respectively. For the case of $H\perp c$, the fitting results in a $C$ of 0.012 emu mol$^{-1}$ Oe$^{-1}$ and a $\theta_W$ of -34 K. For the case of $H||c$, a $C$ of 0.008 emu mol$^{-1}$ Oe$^{-1}$ and a $\theta_W$ of 1 *K* is obtained from the fitting. The $\chi_0$ values for $Sr_3NaIrO_6$ are 6.2×10$^{-4}$ and 5.4×10$^{-4}$ emu mol$^{-1}$ Oe$^{-1}$ for $H\perp c$ and $H||c$, respectively. The $\mu_{eff}$, calculated from the $C$ is about 0.31 $\mu_B$ for $H\perp c$ and 0.28 $\mu_B$ for $H \| c$.

Regarding the $Sr_3AgIrO_6$ single crystals, anisotropic magnetic measurement is not possible due to their small size (~ 300 μm) and crystal morphology. A total weight of ~19 mg $Sr_3AgIrO_6$ single crystals were collected without orientation for magnetic measurement. The $\chi(T)$ curves measured under varied $H$ for $Sr_3AgIrO_6$ single crystals without orientation are shown in Fig. 4(c). Only FC curves are shown because the ZFC and FC curves are overlapping. There is no sign of magnetic order down to 2 K. The $\chi(T)$ data above 50 K can also be fitted with the Curie-Weiss law, resulting in a $\chi_0$ of $3.27\times10^{-4}$ emu mol$^{-1}$ Oe$^{-1}$, $C$ of 0.04 emu mol$^{-1}$ Oe$^{-1}$ and a $\theta_W$ of -35 K. The $\mu_{eff}$ calculated from the $C$ is 0.57 $\mu_B$.

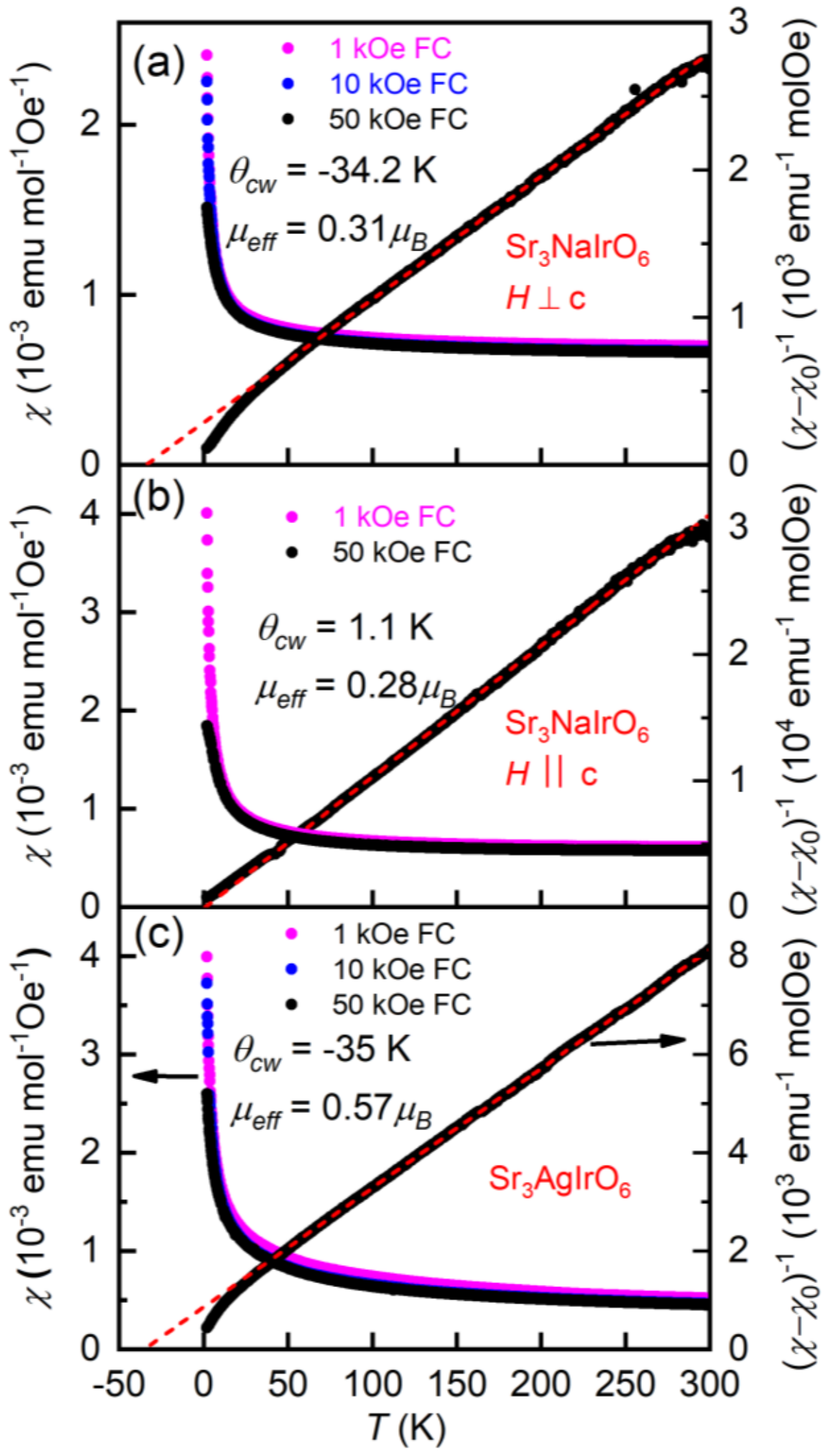

Figure 4. Temperature dependence of magnetic susceptibility and the inverse magnetic

susceptibility for (a) $Sr_3NaIrO_6$ ($H \perp c$), (b) $Sr_3NaIrO_6$ ($H||c$), and (c) $Sr_3AgIrO_6$.

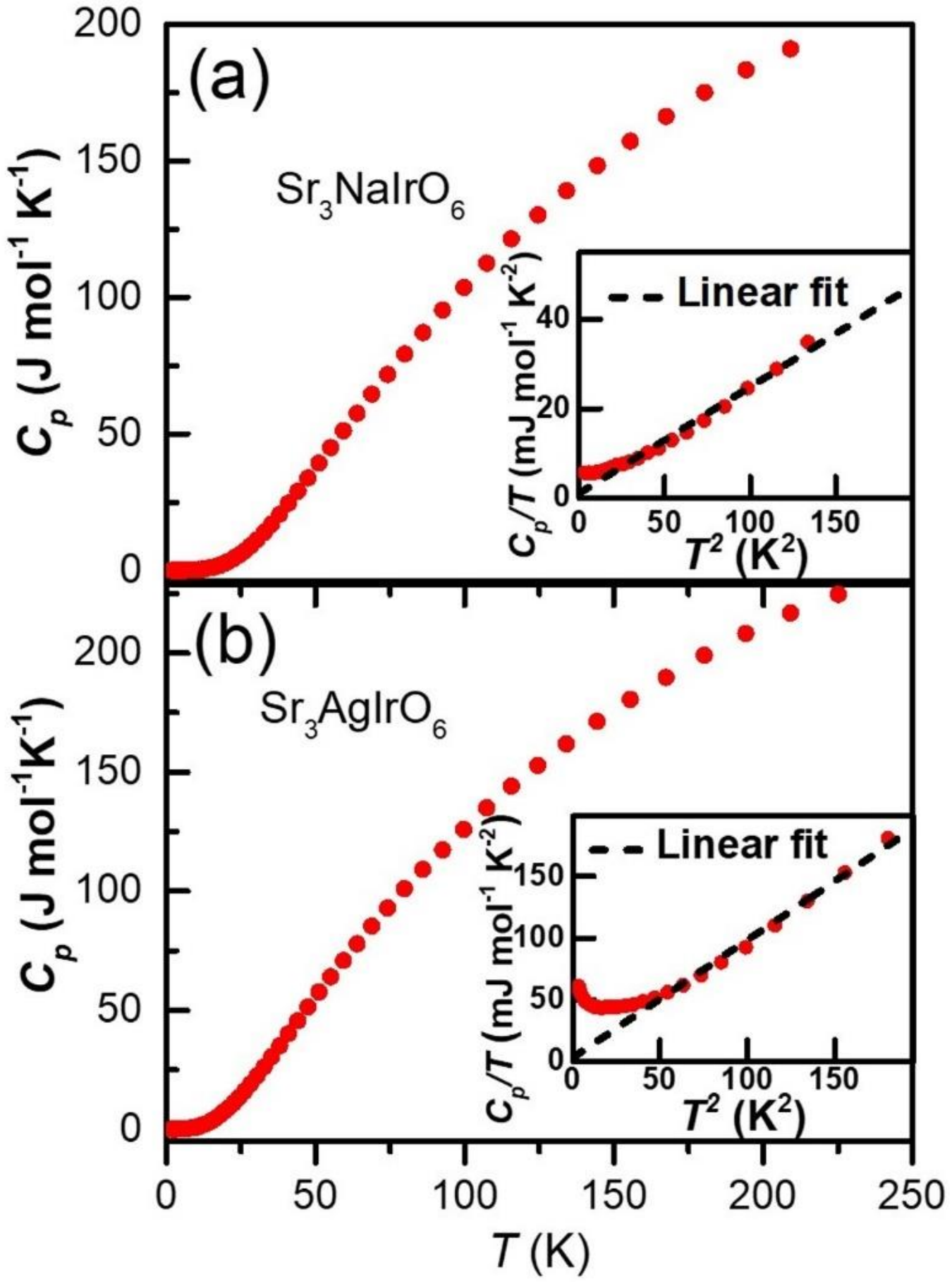

Figure 5. Temperature dependence of specific heat, $C_p$, for (a) $Sr_3NaIrO_6$ and (b) $Sr_3AgIrO_6$. The insets show the corresponding $C_p/T$ vs $T^2$ curves.

Figure 5 shows the $C_p(T)$ curves of $Sr_3NaIrO_6$ and $Sr_3AgIrO_6$. There is no indication of magnetic order down to 1.8 K for $Sr_3NaIrO_6$ and $Sr_3AgIrO_6$. The low-temperature $C_p/T$ vs $T^2$ data, shown in the insets in Figure 5, show rough linear behaviors except for the upturns at the lowest temperatures. The linear parts were analyzed with the approximated Debye model, $C_p/T = \gamma + \beta T^2$, where the $\gamma$ is the electronic specific heat coefficient and the $\beta$ is related to the Debye temperature. Fitting of the linear parts results in a $\gamma = 0.07$ mJ mol$^{-1}$ K$^{-2}$ and $\beta = 2.64\times10^{-4}$ J mol$^{-1}$ K$^{-4}$ for $Sr_3NaIrO_6$ and $\gamma = 3.0$ mJ mol$^{-1}$ K$^{-2}$ and $\beta = 9.05\times10^{-4}$ J mol$^{-1}$ K$^{-4}$ for $Sr_3AgIrO_6$. The small $\gamma$ value indicates the vanishing of the density of states, which is consistent with the insulating nature of $Sr_3NaIrO_6$ and $Sr_3AgIrO_6$.

The low-temperature $C_p/T$ vs. $T$ curves measured without and with varied magnetic fields are shown in Figs. 6(a) and 6(b) for $Sr_3NaIrO_6$ and $Sr_3AgIrO_6$, respectively. The $C_p/T$ vs. $T$ curves measured without fields show small anomalies below ~3.5 K. These anomalies shift towards higher temperatures and broaden with increasing applied magnetic fields, suggesting their magnetic origin. To estimate the magnetic contribution ($C_{mag}$), we subtract the $C_p(T)$ data below 12 K measured under 90 kOe with an estimated lattice contribution $C_{lat} = \gamma T + \beta T^3$, and the resulted $C_{mag}/T$ vs. $T$ curves are shown in Figs. 6(c) and 6(d) for $Sr_3NaIrO_6$ and $Sr_3AgIrO_6$, respectively. These curves show broad maximums around 3 - 4 K for $Sr_3NaIrO_6$ and $Sr_3AgIrO_6$. It should be noted that these values may be under-estimated because the entropy below 1.8 K is not counted in these cases. The estimated magnetic entropy ($S_{mag}$) for $Sr_3NaIrO_6$ and $Sr_3AgIrO_6$ is about 0.06 J $mol^{-1}$ $K^{-1}$ and 0.33 J $mol^{-1}$ $K^{-1}$, respectively. Assuming these magnetic contributions are from $Ir^{4+}$ impurities, these values are about ~1.7% and ~9.2%, respectively, of the averaged entropy $S_{mag}$ = 3.6 J $mol^{-1}$ $K^{-1}$ reported for the $Ir^{4+}$ double perovskite $La_2MgIrO_6$ and $La_2ZnIrO_6$ [36]. These results are consistent with the XAS results that the Ir ions in $Sr_3NaIrO_6$ are $Ir^{5+}$ (but cannot exclude the presence of a few percent of magnetic $Ir^{4+}$ or/and $Ir^{6+}$) while a moderate amount of $Ir^{4+}$ ions is coexisting with the $Ir^{5+}$ ions in $Sr_3AgIrO_6$.

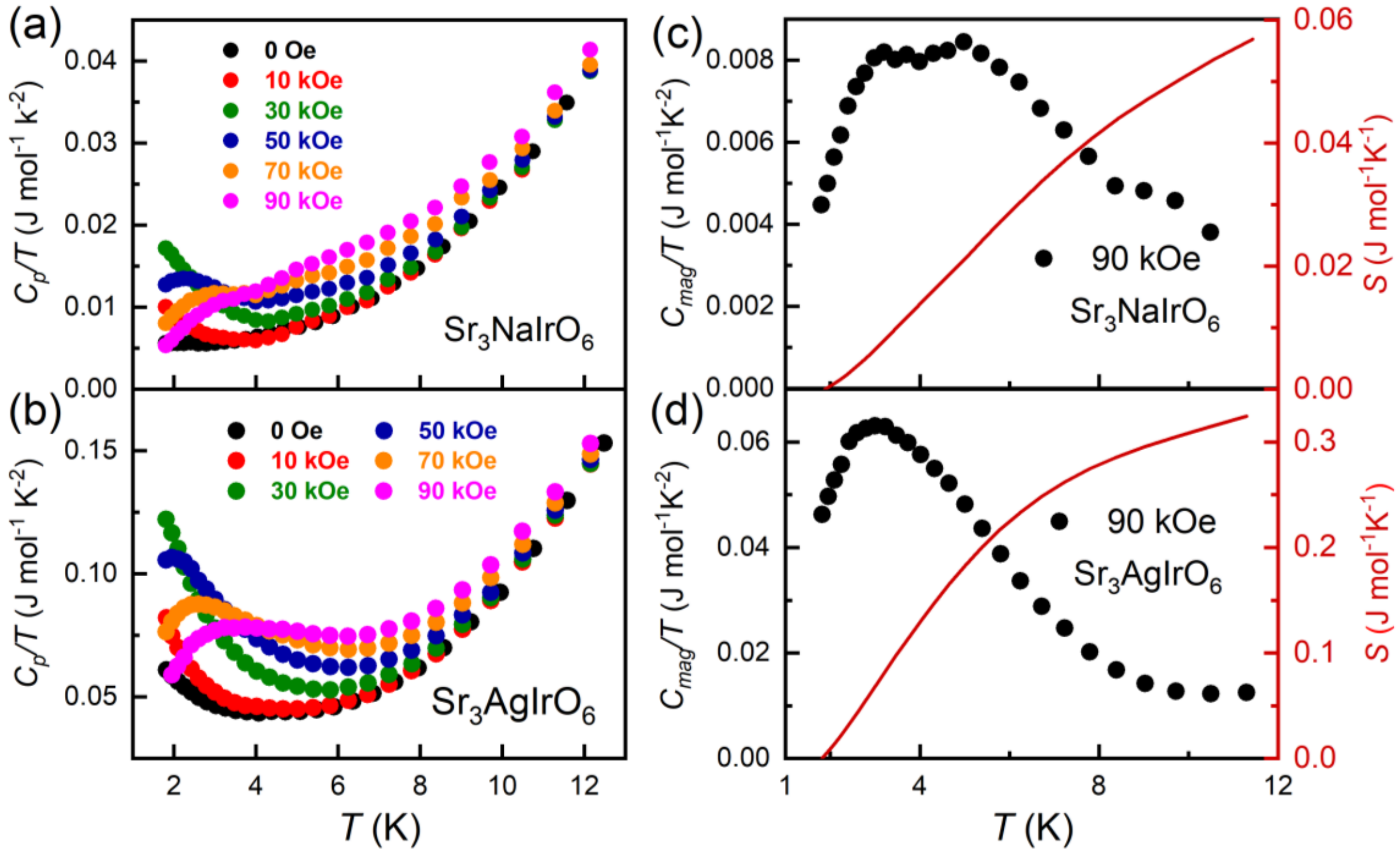

Figure 6. Temperature-dependent $C_p/T$ data under varied magnetic fields for (a) $Sr_3NaIrO_6$ and (b) $Sr_3AgIrO_6$. Temperature-dependent $C_{mag}/T$ and $S_{mag}$ data for (c) $Sr_3NaIrO_6$ and (d) $Sr_3AgIrO_6$.

The $\mu_{eff}$, $\theta_W$, and $\chi_0$ values for $Sr_3NaIrO_6$ and $Sr_3AgIrO_6$ reported in this work are summarized in Table IV in comparison with other $Ir^{5+}$ oxides crystalizing in double perovskite or $K_4CdCl_6$-type structures. The $\chi_0$ summarized in Table IV are of the magnitude of $10^{-4}$ emu mol$^{-1}$ Oe$^{-1}$. The $\chi_0$ reported in this work for $Sr_3NaIrO_6$ and $Sr_3AgIrO_6$ single crystals are in the reported range. Except for the large $\mu_{eff}$ of 0.91 $\mu_B$/Ir and of 1.44 $\mu_B$/Ir once reported for $Sr_2YIrO_6$ [11] and $Ba_2YIrO_6$ [12], respectively, other studies on $Sr_2YIrO_6$, $Ba_2YIrO_6$, and the other $Ir^{5+}$ double perovskite and $K_4CdCl_6$-type oxides reveal relatively small $\mu_{eff}$ values of 0.19 – 0.63 $\mu_B$/Ir [13–25]. The $\mu_{eff}$ values for single crystals of $Sr_3NaIrO_6$ and $Sr_3AgIrO_6$ in this work are within this range.

In comparison with polycrystalline $Sr_3NaIrO_6$ (0.49 $\mu_B$/Ir), the observed $\mu_{eff}$ values for $Sr_3NaIrO_6$ single crystal (0.31 $\mu_B$/Ir for $H \perp c$ and 0.28 $\mu_B$/Ir for $H \| c$) are reduced. The valence state is also confirmed to be mainly $Ir^{5+}$ in polycrystalline $Sr_3NaIrO_6$ [22]. Thus, the reduction of $\mu_{eff}$ for the single crystal samples may be related to the decrease of by-phases and lattice defects because polycrystalline samples synthesized by solid-state reactions are difficult to avoid the minute amount of by-phases and usually have more lattice defects than the single-crystal samples.

Though our XAS spectra confirm that the oxidation state of Ir in $Sr_3NaIrO_6$ is mainly $Ir^{5+}$, but cannot exclude the presence of a few percent of magnetic $Ir^{4+}$ or/and $Ir^{6+}$ ions. Presence of $Ir^{4+}$ and/or $Ir^{6+}$ magnetic defects has been confirmed in $Sr_2YIrO_6$ [20], $Sr_2CoIrO_6$ [35], and $Ba_2YIrO_6$ [18]. Studies on $Ir^{4+}$ double perovskites $La_2ZnIrO_6$ and $La_2MgIrO_6$ have reported $\mu_{eff}$ values of 1.71 and 1.42 $\mu_B$, respectively, which are close to the theoretical value of 1.73 $\mu_B$ for $Ir^{4+}$ ($J = 1/2$) [36]. If we assume the observed $\mu_{eff} \approx 0.3$ $\mu_B$ for $Sr_3NaIrO_6$ single crystals is solely contributed from magnetic $Ir^{4+}$ ($\mu_{eff}$ = 1.73 $\mu_B$) while $Ir^{5+}$ ions are nonmagnetic ($J = 0$), there are should be about ~3.0% of $Ir^{4+}$ magnetic impurities according to $\mu_{eff}^2 = (1-x)(\mu_{eff}\text{-}Ir^{5+})^2+x(\mu_{eff}\text{-}Ir^{4+})^2$. The existence of magnetic $Ir^{4+}$ with a low limit of ~1.7% for $Sr_3NaIrO_6$ is suggested from the analysis of low-temperature $C_p(T)$ data, which is not far from the theoretical amount of ~3.0%.

In comparison with $Sr_3NaIrO_6$, the average valence state of Ir in $Sr_3AgIrO_6$ is a little lower than $Ir^{5+}$, indicating the presence of a moderate amount of $Ir^{4+}$ ions. The presence of magnetic $Ir^{4+}$ with a low limit of ~9.2% is estimated for $Sr_3AgIrO_6$ by the analysis of the low-temperature $C_p(T)$ data. The observed $\mu_{eff}$ for $Sr_3AgIrO_6$ single crystals is ~0.57 $\mu_B$. Assuming the observed $\mu_{eff}$ of 0.57 $\mu_B$ for $Sr_3AgIrO_6$ is solely contributed from magnetic $Ir^{4+}$ while $Ir^{5+}$ ions are nonmagnetic, there should be about ~10.9% of magnetic $Ir^{4+}$ according to $\mu_{eff}^2 = (1-x)(\mu_{eff}\text{-}Ir^{5+})^2+x(\mu_{eff}\text{-}Ir^{4+})^2$, which is comparable to the low limit of $Ir^{4+}$ (~9.2%) estimated from $C_p(T)$ data of $Sr_3AgIrO_6$ single crystals.

In comparison with $A_2YIrO_6$ (A = Sr, Ba), where the $Ir^{5+}O_6$ octahedra are separated by $Y^{3+}O_6$, the $Ir^{5+}O_6$ octahedra are separated by $Na^{1+}O_6$/$Ag^{1+}O_6$ octahedra along the one-dimensional chain in $Sr_3NaIrO_6$ and $Sr_3AgIrO_6$. The much larger charge difference would significantly reduce the antisite disorder between $Na^{1+}$/$Ag^{1+}$ and $Ir^{5+}$ in $Sr_3NaIrO_6$ and $Sr_3AgIrO_6$ as compared with $Sr_2YIrO_6$ and $Ba_2YIrO_6$. Thus, the contributions to the magnetic moment from antisite disorder as discussed for $Sr_2YIrO_6$ and $Ba_2YIrO_6$ should be much reduced in $Sr_3NaIrO_6$ and $Sr_3AgIrO_6$.

From these analyses, the paramagnetic moments observed in $Sr_3NaIrO_6$ and $Sr_3NaIrO_6$ single crystals are likely contributed solely by the magnetic $Ir^{4+}$, supporting the $J$ = 0 ground state for $Ir^{5+}$. This is consistent with the studies on $Ir^{5+}$ double perovskites $A_2BIrO_6$ ($A$ = Ba, Sr; $B$ =Y, Lu, Sc) which support the $J$ = 0 ground state for $Ir^{5+}$ and indicate the magnetic moments are from extrinsic sources [14, 17, 18, 20,21]. Studies on the layered oxide $Sr_2Co_{0.5}Ir_{0.5}O_4$ also support the $J$ = 0 ground state for $Ir^{5+}$ and show that the energy gap between the singlet state and the excited triplet state is large [37]. The presence of magnetic $Ir^{4+}$ indicates that a small number of oxygen vacancies exist in our $Sr_3NaIrO_6$ and $Sr_3AgIrO_6$ single crystals. The possibility of thermodynamic instability of stoichiometric $Ir^{5+}$ oxides and a partial reduction of $Ir^{5+}$ to $Ir^{4+}$ has been suggested by Prof. M. Jansen *et al* [23].

It should be noted that the $\theta_W$ values are anisotropic in our anisotropic magnetic measurements on $Sr_3NaIrO_6$ single crystals. For the polycrystalline $Sr_3NaIrO_6$, the reported $\theta_W$ is -23.6 K. For our $Sr_3NaIrO_6$ single crystals, the $\theta_W$ is about 1 K for the case of $H||c$ but is about -34 K for $H\perp c$, indicating that the magnetic exchange interactions are negligible along the one-dimensional chain ($H||c$) and are mainly for interchains ($H\perp c$). In the ideal $Sr_3NaIrO_6$ structure, there is no direct superexchange Ir-O-Ir path, and the magnetic exchange interactions are mediated by the extended superexchange path Ir-O-O-Ir [38,39]. The nearest Ir-Ir is about 5.78 Å along the chain which is a little bit shorter than the length of 5.89 Å for interchains. But, the nearest O-O distance between nearest $IrO_6$ octahedra along the one-dimensional chain (3.50 Å) is much longer than that of 2.99 Å for interchains, which may responsible for that the magnetic exchange interactions are mainly for interchains in the $Sr_3NaIrO_6$ single crystals.

Table IV. Magnetic properties of $Ir^{5+}$ oxides with double perovskite (DP) and $K_4CdCl_6$ structures.

| Material | Crystal structure | $\chi_0$ ($10^{-4}$ emu mol$^{-1}$ Oe$^{-1}$) | $\mu_{eff}$ ($\mu_B$/Ir) | $\theta_w(K)$ | Reference |
|---|---|---|---|---|---|
| $Ba_2YIrO_6$ | DP | - | 0.3 | -10 | [15] |
| $Ba_2YIrO_6$ | DP | 4.8 | 0.63 | -5 | [16] |
| $Ba_2YIrO_6$ | DP | 5.4 | 0.52 | -4 | [16] |
| $Ba_2YIrO_6$ | DP | 5.4 | 0.50 | -8 | [16] |
| $Ba_2YIrO_6$ | DP | 5.83 | 0.44 | -8.9 | [13] |
| $Ba_2YIrO_6$ | DP | - | 0.31 | - | [17] |
| $Ba_2YIrO_6$ | DP | - | 0.48 | -16 | [18] |
| $Ba_2YIrO_6$ | DP | -3.9 | 1.44 | -149 | [12] |
| $Ba_{1.26}Sr_{0.74}YIrO_6$ | DP | 4.4 | 0.64 | -18 | [12] |
| $Ba_{2-x}Sr_xYIrO_6$ | DP | - | 0.47 | - | [19] |
| $Sr_2YIrO_6$ | DP | - | 0.91 | -229 | [11] |
| $Sr_2YIrO_6$ | DP | 5.90 | 0.21 | -2.8 | [14] |
| $Sr_2YIrO_6$ | DP | - | 0.3 | - | [20] |
| $Sr_{1.6}Ca_{0.4}YIrO_6$ | DP | - | 0.6 | - | [20] |
| $Sr_2LuIrO_6$ | DP | 5.49 | 0.27 | -2.55 | [21] |
| $Ba_2LuIrO_6$ | DP | 4.98 | 0.42 | -13.2 | [21] |
| $Sr_2ScIrO_6$ | DP | 5.43 | 0.32 | -10.7 | [21] |
| $Ba_2ScIrO_6$ | DP | 5.10 | 0.48 | -18.7 | [21] |
| $Bi_2NaIrO_6$ | DP | 6.3 | 0.19 | -6.8 | [23] |
| $LaSrMgIrO_6$ | DP | 3.5 | 0.61 | 7 | [24] |
| $LaSrZnIrO_6$ | DP | 3.9 | 0.46 | 1 | [24] |
| $Sr_3NaIrO_6$ | $K_4CdCl_6$ | - | 0.49 | -23.6 | [22] |
| $Sr_3LiIrO_6$ | $K_4CdCl_6$ | - | 0.4 | -21 | [25] |
| $Sr_3LiIrO_6$ | $K_4CdCl_6$ | - | 0.45 | -71 | [25] |
| $Sr_3NaIrO6$ | $K_4CdCl_6$ | 6.2 | 0.31($H\perp c$) | -34 | This work |
| | | 5.4 | 0.28($H\|\|c$) | 1 | |
| $Sr_3AgIrO6$ | $K_4CdCl_6$ | 3.3 | 0.57 | -35 | This work |

**Conclusions**

Single crystals of $Sr_3NaIrO_6$ and $Sr_3AgIrO_6$ have been successfully grown using hydroxides flux. Analysis of room temperature SCXRD data found that $Sr_3NaIrO_6$ and $Sr_3AgIrO_6$ crystallize in the $K_4CdCl_6$-type structure with the space group $R$-3$c$ (No.167). $Sr_3NaIrO_6$ and $Sr_3AgIrO_6$ are electrically insulating with estimated activation gaps of 0.68 eV and 0.80 eV, respectively. The magnetic results show paramagnetism down to 2 K for both $Sr_3NaIrO_6$ and $Sr_3AgIrO_6$. In comparison with polycrystalline $Sr_3NaIrO_6$ (0.49 $\mu_B$), our $Sr_3NaIrO_6$ single crystals display smaller $\mu_{eff}$ values (0.31 $\mu_B$ for $H\perp c$ and 0.28 $\mu_B$ for $H||c$). The $\mu_{eff}$ for $Sr_3AgIrO_6$ single crystals is about 0.57 $\mu_B$. Combined analyses of the XAS spectra and the low-temperature $C_p(T)$ data indicate that the Ir ions are mainly $Ir^{5+}$ in $Sr_3NaIrO_6$ but there is a low limit of ~1.7% of magnetic $Ir^{4+}$ ions. For $Sr_3AgIrO_6$, the Ir ions are a little lower than $Ir^{5+}$ and it contains magnetic impurity $Ir^{4+}$ with a low limit of ~9.2%. The magnetic impurities are likely to fully explain the observed $\mu_{eff}$ values for $Sr_3NaIrO_6$ and $Sr_3AgIrO_6$ single crystals, supporting the $J = 0$ ground state for $Ir^{5+}$ in $Sr_3NaIrO_6$ and $Sr_3AgIrO_6$.

## Acknowledgments

H.L.F thanks Prof. Liu Hao Tjeng for insightful comments. This work is supported by the Beijing Natural Science Foundation (Z180008), the National Natural Science Foundation of China (12104492, and U2032204), the National Key Research and Development Program of China (2021YFA1400401, and 2017YFA0302900), the Strategic Priority Research Program of the Chinese Academy of Sciences (XDB33010000), the K. C. Wong Education Foundation (GJTD-2018-01), and the Informatization Plan of Chinese Academy of Sciences (CAS-WX2021SF-0102). We also acknowledge support from the Max Planck-POSTECH-Hsinchu Center for Complex Phase Materials.